\documentclass[conference]{IEEEtran}
\usepackage{srcltx}
\usepackage{cite}
\usepackage[final]{graphicx}
\usepackage{amssymb}
\usepackage{psfrag}
\usepackage{subfigure}
\usepackage{url}
\usepackage{amsmath}
\usepackage{array}
\usepackage{epsfig}
\usepackage{latexsym}
\begin{document}

\title{Encoding for the Blackwell Channel with Reinforced Belief Propagation}

\author{
\authorblockN{Alfredo Braunstein}
\authorblockA{Institute for Scientific Interchange \\
Villa Gualino, viale S.Severo 65 \\
10133, Turin, Italy \\
braunstein@isi.it} \and
\authorblockN{Farbod Kayhan}
\authorblockA{
Institute for Scientific Interchange,\\
Politecnico di Torino,\\
10129, Turin, Italy \\
kayhan@isi.it} \and
\authorblockN{Guido Montorsi}
\authorblockA{Politecnico di Torino\\Dipartimento di Elettronica\\
10129, Turin, Italy\\
Email: montorsi@polito.it} \and  
\authorblockN{Riccardo Zecchina} 
\authorblockA{International Center for Theoretical Physics,\\ 
Politecnico di Torino,\\ 
Institute for Scientific Interchange\\
Strada Costiera 11 \\
I-34100, Trieste, Italy \\
Email: zecchina@ictp.it} }

\newtheorem{Theor}{Theorem}[section]
\newtheorem{Lemma}{Lemma}[section]
\newtheorem{Def}{Definition}[section]
\newtheorem{Conjecture}{Conjecture}[section]
\newtheorem{Corollary}{Corollary}[section]

\newcommand{\bin}[2]{
    \left (
        \begin{array}{@{}c@{}}
        #1 \\ #2
        \end{array}
    \right )
}


\maketitle

\begin{abstract}

A key idea in coding for the broadcast channel (BC) is binning, in
which the transmitter encode information by selecting a codeword
from an appropriate bin (the messages are thus the bin indexes).
This selection is normally done by solving an appropriate (possibly
difficult) combinatorial
 problem. Recently it has been shown that binning for the Blackwell
 channel --a particular BC-- can be done by iterative schemes based on Survey
Propagation (SP). This method uses decimation for SP and suffers a
complexity of $\mathcal{O}(n^{2})$. In this paper we propose a new
variation of the Belief Propagation (BP) algorithm, named {\em
Reinforced} BP algorithm, that turns BP into a solver. Our
simulations show that this new algorithm has complexity
$\mathcal{O}(n \log n)$. Using this new algorithm together with a
non-linear coding scheme, we can efficiently achieve rates close
to the border of the capacity region of the Blackwell channel.
\end{abstract}

\IEEEpeerreviewmaketitle

\section{Introduction}
\label{SEC:INTR} 
\begin{figure}
    \centering
    \epsfig{file=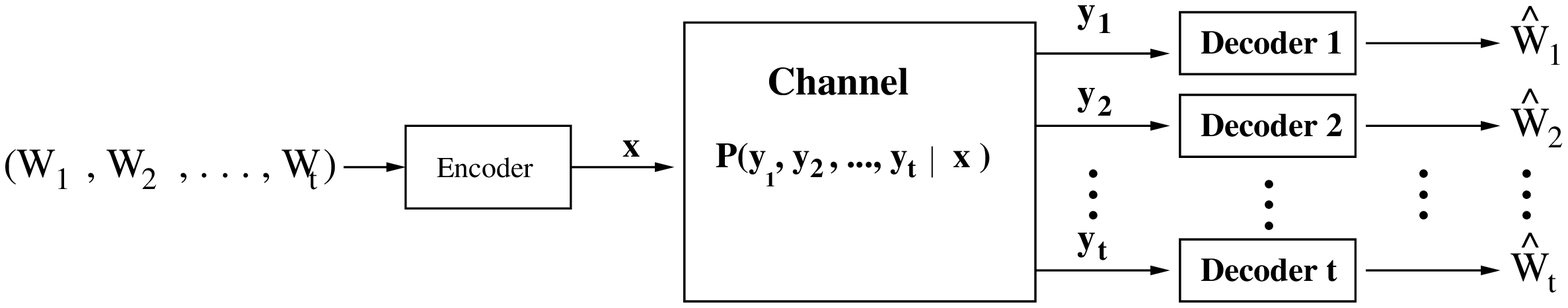,width=0.48\textwidth}
    \caption {\label{t-receiverBC} A single sender and $t$ receivers broadcast channel.}
\end{figure}

Broadcast channels (BC) were first introduced and
analyzed by Cover \cite{Cover}. The general BC with $t$ receivers is
depicted in Fig. \ref{t-receiverBC}. In a BC, a single transmitter
sends simultaneously independent information to multiple receivers.

Coding for each receiver independently with a normal point-to-point
code and sending the $t$ messages sequentially -by allocating
proportions of time to each receiver- is known as time sharing
strategy. It is shown in \cite{Cover} that jointly optimized codes
can have a larger capacity region for error--free communication than
that of time sharing codes ~\cite{Bergmans,Cover,CoverBook}.

A key idea in coding for the BC is the binning strategy, which allows
the transmitter to encode information by selecting a codeword from an
appropriate bin. In this paper we deal with practical implementation
of random binning for the BC. Existing practical binning schemes for
BC are often based on structured codes and maximum likelihood
algorithms. Martinian and Yedidia in \cite{MartinianYedidia} have used
for the first time the random codes on graphs for quantization of a
binary erasure source. Still their method works only for erasure
sources and is not applicable to the general BC.

Recently, Wei Yu and M. Aleksic \cite{WeiYu} showed that the
binning problem for a particular BC, namely the Blackwell Channel
(BWC), when coding is performed by random low-density parity-check
like codes, can be thought as a constraint satisfaction problem.
They proposed an iterative {\em encoder} that works well at rates
close to the border of the BC capacity region.

The main difference of this problem with that of decoding classical
codes is that this combinatorial problem admits many solutions. In
fact in these cases the application of BP allows to compute the
cardinality of the solution space but fail to find a particular
solution.

In \cite{WeiYu} they use Survey Propagation (SP) algorithm for
encoding, fixing one variable after each convergence (decimation).
The main drawback of this method is the encoding complexity which
grows as $\mathcal{O}(n^2)$. Also the decimation works well only when
the connectivity of XOR nodes are very small ($c = $ 2, 3 and 4).

In this paper we use a modified version of BP, called
\emph{Reinforced Belief Propagation} (RBP), originally proposed in
the context of perceptron learning~\cite{AlfredoNN}, which
effectively turns BP into a solver. Experiments show that RBP does
not converge for factor graphs with XOR function nodes. To
overcome this, we propose a new class of sparse non-linear codes.
These two modifications result in a more efficient encoding
complexity (from $\mathcal{O}(n^2)$ to $\mathcal{O}(n\log n )$)
and a lower \emph{Frame Error Rate} (FER), i.e., the probability
of not finding a solution to the encoding problem.

This paper is organized as follows. In the next section we introduce
the general framework of broadcast channels and their capacity
regions. In section \ref{SEC:RBP} we present the iterative updates for
BP and RBP algorithms. Our scheme for coding for the BWC using
non-linear nodes is explained in section \ref{SEC:DEC}. Our results
are presented in section \ref{SEC:RESULT}. The final section is
devoted to conclusions and outlooks.

\section{Notations and Basic Concepts}
\label{SEC:CRBC} In this section we first introduce the basic
concepts and then briefly review some results on capacity region
for deterministic broadcast channels.

\begin{Def}
\label{DEF:BC} A broadcast channel consists of an input alphabet
$\mathcal{X}$, two output alphabets $\mathcal{Y}_1$ and
$\mathcal{Y}_2$ and a probability transition function
$P(\mathbf{y}_1,\mathbf{y}_2|\mathbf{x}).$ The channel is said to be
memoryless if $$P(\mathbf{y}_1,\mathbf{y}_2|\mathbf{x})=
\prod_{i=0}^{n} P(y_{1i},y_{2i}|x_i).$$
\end{Def}

A $\big( (2^{nR_1},2^{nR_2}),n \big)$ code for a BC with independent
information consists of an encoder
$$\mathcal{E}: 2^{nR_1} \times 2^{nR_2} \rightarrow \mathcal{X}^{n}, $$
and two decoders
$$\mathcal{D}_1 : \mathcal{Y}_{1}^{n} \rightarrow 2^{nR_1} \; \;, \; \; \; \;
\mathcal{D}_2 : \mathcal{Y}_{2}^{n} \rightarrow 2^{nR_2}.$$

We assume that the transmitted message pair $(W_1,W_2)$ is uniformly
distributed over the set $2^{nR_1} \times 2^{nR_2}$. The probability
of error $P_{e}^{n}$ is defined to be
$$P_{e}^{n} = P (W_1 \neq  \hat{W}_1 \;\;\; or \;\;\; W_2 \neq \hat{W}_2).$$

\begin{Def}[Capacity Region] A rate pair $(R_1,R_2)$ is called
achievable for the BC if there is a sequence of $\big\{ \big(
(2^{nR_1},2^{nR_2}),n \big) \big\}_{n}$ codes with $P_{e}^{n}
\rightarrow 0$ as $n \rightarrow \infty$. The capacity region of the
broadcast channel is the closure of the set of achievable rates.
\end{Def}

A broadcast channel is deterministic if the channel transition
probabilities are deterministic, i.e.,
$P(\mathbf{y}_1,\mathbf{y}_2|\mathbf{x})$ is a $0-1$ function. The
largest achievable rate region for a general BC using the binning
strategy is known as the Marton's region \cite{Marton}. This region
is proved to be the capacity region for a discrete deterministic
channels \cite{MartonDeterministic}.

\begin{figure}
    \centering
    \epsfig{file=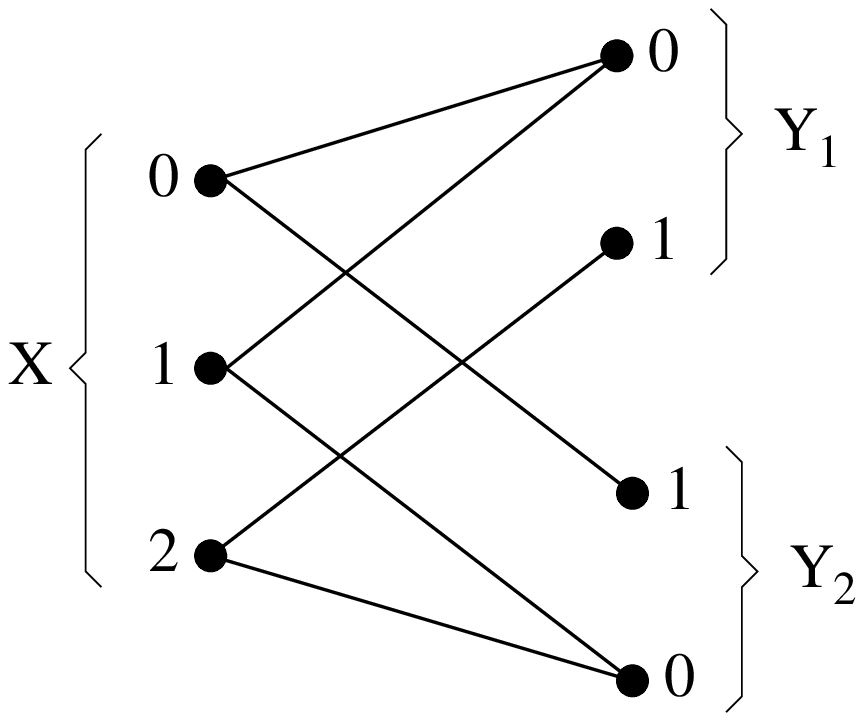,width=0.18\textwidth}
    \caption {\label{Blackwell} The Blackwell channel.}
\end{figure}

A well-known example of a deterministic BC is the BWC (see Fig.
\ref{Blackwell}). The BWC has one input with three symbols and two
outputs each one with two symbols. Given two messages $W_1$ and
$W_2$, the goal is to find the codewords $\mathbf{y}_1 \in 2^{nR_1}$
and $\mathbf{y}_2 \in 2^{nR_2}$ such that $(y_{1i},y_{2i}) \neq
(1,1)$ for $i=1,2,...,n$. The other three combinations are allowed
and they can be reached by selecting one of the three input symbols
of the channel. Even though this channel is not realistic, it is a
non-trivial BC which illustrates the conflict between transmitting
information to first receiver and transmitting to second receiver
\cite{Gelfand}.

\begin{figure}
    \centering
    \epsfig{file=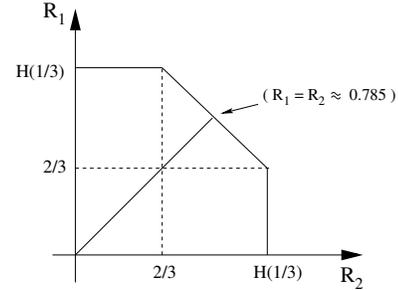,width=0.28\textwidth}
    \caption {\label{CapaRegion} Rate region for the BWC with uniform distribution.}
\end{figure}

Since the channel is deterministic we have $H(X) = H(Y_1,Y_2)$. In the
rest we assume a uniform probability distribution over $X$.
With this input distribution the capacity region for BWC becomes

\begin{eqnarray*}
R_1 & \leq  & H(\frac{1}{3})\\
R_2  & \leq  & H(\frac{1}{3})\\
R_1 + R_2 & \leq & \log_{2} 3.
\end{eqnarray*}
This capacity region is shown in Fig.~\ref{CapaRegion}.

\section{BP and RBP algorithms}
\label{SEC:RBP}

Let $g: S \subset \mathbb{R}^{n} \rightarrow \mathbb{R}$ be a real
valued function
over the domain $S$ and
\begin{equation}
    \label{FG}
    g(x_1, x_2,...,x_n) \propto \prod_{j \in M} f_j(X_j)
\end{equation}

where $X_j$ is a subset of the set of variables.

\begin{Def} A factor graph of a function $g$ factorized as in (\ref{FG}) is a bipartite graph
    with $n$ vertex in one part (variable nodes) and $M$ vertex in the
    second part (factor nodes). An edge connects variable node $x_i$
    to factor node $f_j$ if and only if $x_i$ is an argument of the
    local function $f_j$, i.e., $x_i \in X_j$.
\end{Def}

We show the $i$th marginal function associated with $g(x_1,
x_2,...,x_n)$ by
$$g_i (x_i) \propto \sum_{\sim\{ x_i \}} g(x_1, x_2,...,x_n)$$
where the symbol $\sim\{ x_i \}$ indicates the set of all variable
configurations with the $i$-th variable fixed to $x_i$.

Calculating the marginal functions in general is a hard task. BP is
an efficient and exact algorithm to calculate all marginal functions
$g_i(x_i)$ when the factor graph of $g$ is cycle-free. It is
possible to use BP also in the presence of loops. The resulting
algorithm will be iterative and  calculates the marginals
approximately. In the rest of this section, first we review the BP
update rules and then present a generalization of BP called
\emph{Reinforced BP algorithm} (RBP) \cite{Chavas}.

Let $\mu_{x \rightarrow f}^{\ell}(x)$ denotes the message sent
form variable node $x$ to factor node $f$ at the $\ell$th
iteration. Similarly, $\mu_{f \rightarrow x}^{\ell}(x)$ denotes
the message sent from factor node $f$ to variable node $x$ at the
iteration $\ell$. Also, let
\begin{eqnarray*}
    \mathcal{N}(x_i) &\triangleq & \{ j | x_i \in X_j \},\\
    \mathcal{M}(f_j) &\triangleq & \{ i | x_i \in X_j \},
\end{eqnarray*}
then the BP algorithm messages can be expressed as
follows:

{\bf Local Function to Variable:}
\begin{equation}\label{SPfunc2var}
    \mu_{f_j \rightarrow x_i}^{\ell}(x_i) \propto \sum_{\sim \{x_i\}} \Big( f_j(X_j)
    \prod_{l \in \mathcal{M}(f_j) \setminus \{i\}} \mu_{x_l \rightarrow
    f_j}^{\ell}(x_l)\Big)
\end{equation}

{\bf Variable to Local Function:}
\begin{equation}\label{SPvar2func}
    \mu_{x_i \rightarrow f_j}^{\ell+1}(x_i) \propto \prod_{l \in \mathcal{N}(x_i) \setminus
    \{j\}} \mu_{f_l \rightarrow x_i}^{\ell}(x_i)
\end{equation}

For $\ell=1$, we initialize the messages $\mu_{x \rightarrow
f}^{\ell}(x)$ randomly. These updating rules tell us how to produce
locally outgoing messages from incoming messages. We define the
marginal function of variable $x_i$ at iteration $\ell+1$ as

\begin{equation}\label{LocalField}
    g_{i}^{\ell+1}(x_i) \propto \prod_{l \in \mathcal{N}(x_i)} \mu_{f_l \rightarrow
    x_i}^{\ell}(x_i).
\end{equation}

\begin{figure}
  \centering
  \epsfig{file=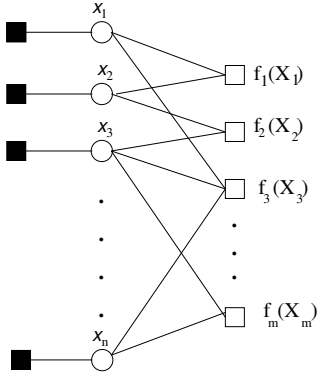,width=0.23\textwidth}
  \caption {\label{Fig:RBPfactor} The modified factor graph for
    RBP. The black squares are dynamic nodes which their value is a
    function of the marginal of the related variable at a preceding
    iteration.}
\end{figure}

The algorithm converges after $t$ iterations if and only if for all
variables $x_i$ and all function nodes $f_j$
$$ \mu_{f_j \rightarrow x_i}^{t+1}(x_i) = \mu_{f_j \rightarrow
x_i}^{t}(x_i) $$
In practice we need to predefine maximum number of iterations
$\ell_{\max}$ and a precision parameter $\epsilon$ as the input to the
algorithm.

BP has been generalized/modified in many ways~\cite{Yedidia, Ricca,
Chavas,BPbased,TAPmurayama}. BP and its generalizations have proven to be 
efficient when the variables are biased toward a solution. 
Unfortunately when this condition is not 
fulfilled marginal 
themselves are not sufficient to find a solution to the combinatorial 
problem and one has to resort to some decimation techniques (\cite{Ricca}, \cite{WeiYu}), 
resulting in a high computational complexity. 

We will show here the RBP
equations~\cite{AlfredoNN} that turn BP into an efficient solver.
The idea is to introduce a new set of reinforcement messages which
drive the equations toward a single solution. First we modify the
original factor graph by adding to each variable node a new function
node. In Fig. \ref{Fig:RBPfactor} these new function nodes are
depicted by black squares. These function nodes are dynamic and at
the $\ell$th iteration take the value $ \big( g_{i}^{\ell-1}(x_i)
\big) ^{\gamma(\ell-1)}$, i.e, a power of the marginal of the
variable $x_i$ at the preceding iteration. $\gamma(\ell)$ is a non
decreasing function in $[0,1]$ with $\gamma(0) = 0$. While the
updating rule (\ref{SPfunc2var}) at function nodes does not change
for RBP, the variable to function messages should be modified as
below.

{\bf Variable to Local Function for RBP:}
\begin{equation}\label{RBPvar2func}
    \mu_{x_i \rightarrow f_j}^{\ell+1}(x_i) \propto  \big(g_{i}^{\ell}(x_i) \big)^{\gamma(\ell)}
    \prod_{l \in \mathcal{N}(x_i) \setminus
    \{j\}} \mu_{f_l \rightarrow x_i}^{\ell}(x_i).
\end{equation}

In this paper we deal only with binary constraint satisfaction
problems, where $\mathbf{x} \in \{ 0,1 \}^{n}$ and the local
functions $f_{j}(X_j)$ are 0-1 indicator functions.  A vector
$(x_1,x_2,...,x_n)$ satisfies $f_j(X_j)$ if $f_j(X_j) = 1 $.
$(x_1,x_2,...,x_n)$ is called a solution of the constraint
satisfaction problem if all local functions are satisfied, i.e.,
$\prod_{j \in M} f_j(X_j) = 1$. It is easy to show that if RBP
converges, it converges to a solution of our problem (all messages
completely polarized to delta functions). This simple modification
provides us with a solver with complexity $\mathcal{O}(n)$ (assuming
roughly constant convergence time). Note that the number of iteration
of RBP depends also on the choice of $\gamma(\ell)$ in
(\ref{RBPvar2func}). As our experiments show, choosing an optimal
$\gamma$ can dramatically decrease the number of iterations of RBP at
least for the binning problem. For the rest of this paper we will set
\begin{equation}
\label{coolingfunction} \gamma(\ell) = 1 -
\gamma_{0}\gamma_{1}^{\ell},
\end{equation}
where $\gamma_{0},\gamma_{1}$ are in $[0,1]$.

\section{Coding for the Blackwell Channel Using Non-Linear Nodes}
\label{SEC:DEC} One of the main coding strategies for deterministic
broadcast channel is binning. The idea is to generate $2^{nH(Y_1)}$
codewords $\mathbf{y}_1$ and $2^{nH(Y_2)}$ codewords $\mathbf{y}_2$
and randomly assign them into $2^{nR_1}$ and $2^{nR_2}$ bins. To
transmit a particular pair of bin indices $(i,j)$, the transmitter
looks for a pair of codeword $(\mathbf{y_1},\mathbf{y_2}) \in (i,j)$
such that they are jointly typical.

For the BWC, the joint typicality of $\mathbf{y_1}$ and
$\mathbf{y_2}$ is equivalent to being consistent with the channel
constraints. Therefore, we are looking for efficient ways to finding
a pair $(\mathbf{y_1},\mathbf{y_2})$ such that $(y_{1i},y_{2i}) =
(1,1)$ does not occur for $i=1,2,...,n$.

Wei Yu and Marko Aleksic in \cite{WeiYu} have suggested a random
binning method for BWC based on low-density parity-check like codes.
In this section we first review their results and then modify their
scheme using non-linear nodes and RBP algorithm. As we will see in
the next section, these modifications imply a better encoding
complexity and a lower FER for large function node connectivity.

\begin{figure}
    \centering
    \epsfig{file=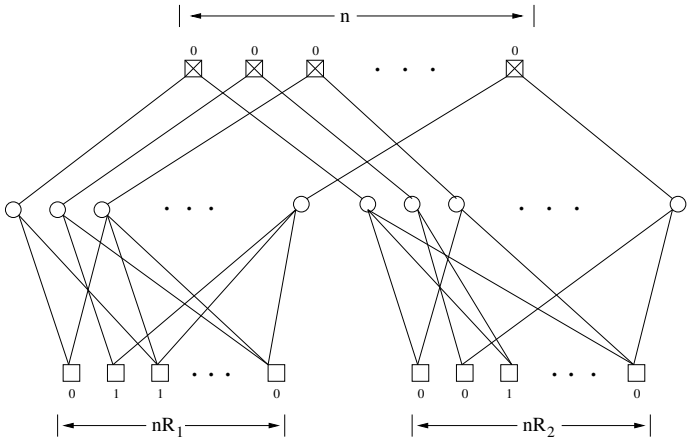,width=0.5\textwidth}
    \caption {\label{codeBlack} Factor graph for LDPC like encoding for Blackwell channel.}
\end{figure}

Fig. \ref{codeBlack} illustrates the graphical structure (factor
graph) of the coding scheme used in \cite{WeiYu}. $n$ circles
denote the variable nodes, $nR_1+nR_2$ squares denote the parity
check nodes and $n$ crossed squares denote the product constraints
(ensuring the $(1,1)$ pair does not occur).
The encoding process is as follows.
The information bits $nR_1$ and $nR_2$ are placed at the parity
checks. These values are actually the bin indices. The goal is to
find the set of variable assignments that satisfy the
parity-checks and product constraints simultaneously. These two
sets of constrains ensure the typicality of the pair
$(\mathbf{y_1},\mathbf{y_2})$. When $n$ is large an exhaustive
search is not feasible and practical algorithms are desirable.

In~\cite{WeiYu} the survey propagation algorithm is suggested for
this encoding problem.
The main drawback of using SP/BP is the complexity which grows as
$\mathcal{O}(n^2)$ because of the decimation process. As it was
also mentioned in \cite{WeiYu}, this method works only for small
function nodes connectivity.

On the other hand, the RBP algorithm, introduced in the section
\ref{SEC:RBP}, do not converge --even for rates not close to the
capacity-- for linear codes. To overcome this, we substitute parity
check nodes with non-linear (random) functions. These kind of gates 
have been analyzed with methods from
statistical physics \cite{CilibMezard}. Intuitively, the reason for which 
random gates may show a better performance with respect to the linear nodes 
can be explained as follows. 
Strong symmetry
properties of XOR functions do not allow a decimation procedure to
choose a good decimation path that preserves the
uncorrelation hypothesis needed for BP; indeed, in any
decimation step with XOR gates, undecided variables have all equal
probability of taking 0 or 1.
 
Given $c$ variable input nodes we choose a
non-linear function node randomly from all $2^c$ possible balanced
truth-tables. We eliminate from this choice fully-canalized nodes,
i.e., random nodes for which a particular value of one of their
variables determine the output. For our code constructions we have
used 4 to 8 different random nodes for each connectivity $c$. 
Note that the complexity of updating messages on a random node with 
degree $c$ is of order $2^c$. In this paper we confine ourselves
to a constant degree $c=6$ and hence ignore this factor in the rest.

In order to show the suitability of non-linear nodes to the
problem at hand we compute the normalized size of the solution
space, defined as $H_S=\log(N_s)/n$, where $N_s$ is the number of
solutions. An approximation to $H_S$ can be computed directly from
the BP messages at a fixed point \cite{Yedidia}.

\begin{figure}
    \centering
    \epsfig{file=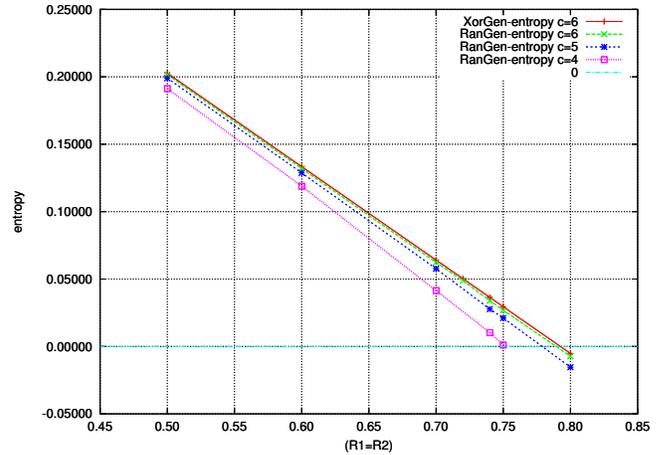,width=0.48\textwidth}
    \caption {\label{Fig:Entropy} Entropy as a function of rate ($R_1=R_2$) for different
    function node connectivity $c$. At any given rate
    the entropy of codes with non-linear
    factor nodes increases with $c$ and approaches the entropy of
    linear codes.}
\end{figure}

In Fig. \ref{Fig:Entropy} we plot $H_S$  as a function of rate using
linear nodes and non linear nodes for different values of the
function node connectivity $c$. The entropy of codes with non-linear
function nodes increases with $c$ and approaches the entropy of
linear codes. Note that for linear codes the entropy does not change
with connectivity.
A connectivity $c=6$ thus guarantees a solution space with
cardinality near to those of LDPC codes when using non-linear type
nodes. This value of connectivity has then been chosen for the
code construction.


\section{Results}
\label{SEC:RESULT} Table \ref{Tab:FER} shows the FER and BER of our
constructed non-linear codes for the BWC with $n=1000$ and constant
connectivity $c=6$ at different rates. The last line reports the
values we chose for $\gamma_1$.
\begin{table}
\centering
\begin{tabular}{|r|c|c|c|c|c|c|l|}
\hline
Rate & 0.5 & 0.6 & 0.7 & 0.72 & 0.73 & 0.74 & 0.75\\
\hline
FER & 0 & 0 & 0.03 & 0.1 & 0.35 & 0.825 & 0.975\\
\hline
BER & 0 & 0 & 0.00011 & 0.0013 & 0.00425 & 0.0119 & 0.0347\\
\hline

$\gamma_{1}$ & 0.99 & 0.995 & 0.999 & 0.9995 & 0.9999 & 0.99999 & 0.999995\\
\hline
\end{tabular}
\caption{\label{Tab:FER} BER and FER of non-linear LDPC like encoders
at a given rate ($R_1 = R_2$) and connectivity $c=6$.}
\end{table}

We estimated the algorithmic complexity of the presented coding
scheme in a series of experiments described below. In particular, we
will show how the convergence time changes as a function of $n$ and
$\gamma_{1}$ . The RBP algorithm was run with an estimated optimal
value of $\gamma_{1}$, and we have chosen a cutoff time of
$\frac{1}{(1-\gamma_{1})}$ to measure the bit and frame error rates.

\begin{figure}
    \centering
    \epsfig{file=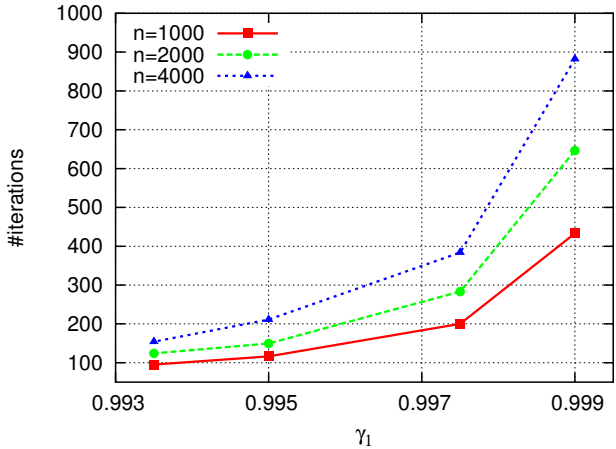,width=0.48\textwidth}
    \caption {\label{Fig:numofit} The average number of needed iterations as a
    function of $\gamma_{1}$ at rate $R_1=R_2=0.70$ for different values of $n$.  Note that
    for smaller $\gamma_{1}$ we need less number of
    iterations but both BER and FER are larger (see Fig. \ref{Fig:BER} and table \ref{Tab:FER}). }
\end{figure}

\begin{figure}
    \centering
    \epsfig{file=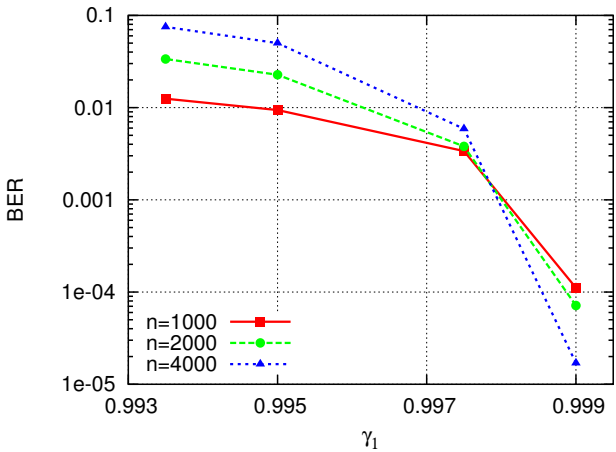,width=0.48\textwidth}
    \caption {\label{Fig:BER} Bit error rate as a function of  $\gamma_{1}$ at rate $R_1=R_2=0.70$ for
    different values of $n$. Error bars are smaller than symbols size in this scale. }
\end{figure}

Fig. \ref{Fig:numofit} shows the average number of iterations needed
(in the case of success) for a rate $R_1=R_2=0.70$ as a function of
$n$ and $\gamma_{1}$ and for 160 encoding operations. These
simulations indicate that the number of iterations increase as
$\mathcal{O}(\log n)$. Although the number of iterations increase
(exponentially) with $\gamma_{1}$, both the BER and FER decrease
(exponentially) as it can be seen in Fig.~\ref{Fig:BER}. Note that
for rates closer to the capacity bound ($ R_1=R_2 \approx  0.785$) a
value of $\gamma_{1}$ closer to $1$ (and larger number of
iterations) is needed.

Although the results depicted in Fig. \ref{Fig:numofit} indicate a
logarithmic increase in the number of iterations as a function of
$n$, this result may be due to a not optimized choice of
$\gamma(\ell)$. For example by choosing $\gamma_{0} = 0.8$ in
(\ref{coolingfunction}) it is possible to reduce the number of
iterations for $n=4000$ and $\gamma_{1}=0.999$ by nearly $25\%$.
In other words, one can avoid approximately the first $200$
iterations of RBP without loosing in performance.

\section{Conclusion and Outlooks}
\label{SEC:CONC} We have introduced a novel variation of the BP
algorithm, called reinforced BP, that turns it into an efficient
solver for non-linear problems even when they have a large
solution space. The algorithm have the same complexity of BP and
thus considerably smaller than the decimation approach applied to
BP/SP proposed in \cite{WeiYu}.

Using RBP we have constructed a general and rather efficient encoding
scheme for the BWC. Our codes provide good encoding performances for
rates up to $0.72$. This result can be possibly improved by
optimizing the function $\gamma(\ell)$ and the degree distributions
of the code.

Our scheme compares well with existing ones: for linear codes with $R
= R_1 = R_2 = 0.75$ and decimation, as it was reported also in
\cite{WeiYu}, one can get the bit error rate of $5.10^{-3}$. Still,
simulations show that the FER in this case is 0.9 and it does not
improve for smaller rates like $R=0.72$ with the same connectivity.
On the other hand it works only for low function node connectivity.
Our scheme is much more flexible and provides a comparable FER and
BER at $R=0.75$ with lower computational complexity. For smaller
rates our codes outperform the existing linear encoding schemes.


\end{document}